\documentclass[preprint,aps]{revtex4}

\begin{document}


\title{Pure decoherence in quantum systems}

\author{Robert Alicki}

\address{Institute of Theoretical Physics and Astrophysics, University
of Gda\'nsk, Wita Stwosza 57, PL 80-952 Gda\'nsk, Poland}

\date{\today}

\begin{abstract}
A popular model of decoherence based on the linear coupling
to harmonic oscillator heat baths is analized and shown to be
inappropriate in the regime where decoherence dominates over energy dissipation,
called pure decoherence regime.
The similar mechanism essentially related to the energy conservation implies that,
on the contrary to the recent conjectures, chaotic environments can be less efficient decoherers than regular ones. Finally, the elastic
scattering mechanism is advocated as the simplest source of pure decoherence.  
\end{abstract}


\maketitle
Decoherence became one of the most popular topics in the physical literature of the
last decade [1-4]. This is mainly due to the progress in experimental techniques
allowing to observe the onset of decoherence at the most interesting regime i.e.
at the border between quantum and classical worlds [5]. Another motivation is 
a destructive role
of decoherence in the possible future technology based on
quantum information processing [6]. Despite the fact that the theoretical models of 
decoherence exist at least for 40 years [7] a closer look at certain aspects of these theories 
reveals inconsistencies and interpretational problems. 
We shall use the following definition: 
{\it Decoherence is an irreversible, uncontrollable and persistent formation of
quantum correlations (entanglement) of the system with its environment.}
\par
Usually, decoherence is accompanied by {\it dissipation} i.e. the net exchange of energy
with environment. For the sake of clarity we shall restrict ourselves to the case
of {\it pure decoherence} called also {\it dephasing} for which the process of energy
dissipation is neglible. This situation occurs in two cases: A) for system's Hamiltonian
$H_S$ commuting (approximatively) with the system-bath interaction Hamiltonian $H_{int}$;
B) for the initial states of $S$ which evolve very slowly under the dynamics governed by $H_S$
on the time scale of decoherence processes. In both cases we can disregard the presence of system's Hamiltonian $H_S$ in the derivations concerning decoherence processes. As an example of
B) we can consider a system equivalent to a 1-dimensional particle in a symmetric double-well
potential coupled to a bath. In the semiclassical regime we can rectrict ourselves to the  2-dimensional Hilbert space of initial states spanned by the lowest lying almost degenerated Hamiltonian eigenstates [8]. The second example is a heavy particle interacting with a medium. As initial states for its center of motion we can choose superpositions of well-localized wave packets with
small enough kinetic energies. For decoherence effects grow with the size of the particle
while the kinetic energy exchange decreases we are again in the pure decoherence regime.
\par
 Pure decoherence is supposed to be the main ingredient of the theory
explaining the apparent absence of superpositions of macroscopically distinguishable states
and the transition from quantum to classical world. Indeed, the explanation of the rapid
decay of quantum correlations between "Schr\"odinger cat" states should not essentially depend 
on the energy difference between $|dead\ cat>$ and $|alive\ cat>$ but is rather related
to the distinguishability of these states described in terms of certain collective observables
which are coupled to the environment.
\par
In this letter we show that the most popular model of quantum open system  based on the linear coupling to the harmonic oscillator bath is inadequate in the pure decoherence regime. Using
a unifying approach in terms of reservoir's spectral function we also show that, contrary to the recent conjecture, chaotic environments can be less efficient decoherers than regular ones  at least for pure decoherence case.
\par
The first problem can be understood in simple physical terms.
Namely, pure decoherence in the open system must be accompanied by the irreversible perturbation
of the environment's state but the energy of the environment should be asymptotically preserved.
However, the linear coupling to the bosonic environment implies that the only
change of its state is caused by irreversible processes of emission and absorption
of single bosons which must alter the environment's energy. The energy exchange can be reduced by 
a strong coupling to low energy bosons what in turn produces infrared divergencies. Those divergencies change completely the physical interpretation of the model, in particular the decomposition of the total system into the open system $S$ and the reservoir $R$, what seems
not to be taken into account in the literature [9,10].  
\par
{\it Spin-boson model} [11]. To explain this effect in a rigorous way we consider a two level system coupled linearly to the bosonic reservoir that is defined in terms of fields satisfying canonical commutation relations 
$[a(\omega), a^{\dagger}(\omega')]= \delta (\omega - \omega')$,
a single-boson Hilbert space $L^2[0,\infty)$, a single-boson Hamiltonian $h_1$,
$(h_1f)(\omega)= \omega f(\omega)$ and the second quantization Hamiltonian$$
H_B =  \int_0^{\infty} d\omega\,\omega\,a^{\dagger}(\omega)a(\omega)
\eqno(1)
$$
acting on the bosonic Fock space ${\cal F}_B \bigl(L^2[0,\infty)\bigr)$ with the vacuum
state $\Omega$. The general spin-boson Hamiltonian depending on the 
function ("formfactor") $g(\omega)$ can be written as ($\sigma_k$ - Pauli matrices)
$$
{\bf H}_g = H_S + H_B + H_{int}(g)\ ,\ \ H_S = {1\over 2}\epsilon\sigma_1 
\eqno(2)
$$ 
where the interaction Hamiltonian is linear in bosonic field and reads
$$
H_{int}(g) = \sigma_3\otimes \int_0^{\infty}d\omega \,\omega\bigl( {\bar g}(\omega) a(\omega)
+ g(\omega) a^{\dagger}(\omega)\bigr)\ .
\eqno(3)
$$
The Hamiltonians act on the Hilbert space 
${\cal H}_{SB} = {\bf C}^2\otimes {\cal F}_B \bigl(L^2[0,\infty)\bigr)$.
The system Hamiltonian $H_S$ describes the coherent tunneling between two eigenstates of 
the "position" operator $\sigma_3\psi_{\pm} = \pm\psi_{\pm}$. 
We assume that this process is much slower then the
decoherence i.e. $\hbar/\epsilon >> \tau_{dec}$ . This is precisely the condition for pure 
decoherence
regime in our model which allows to put $\epsilon = 0 $ in (3) and leads to

$$
{\bf H}_g =\pmatrix{H_{+g} & 0      \cr
                     0  & H_{-g} \cr}\  \equiv {\rm diag}[H_{+g}, H_{-g} ]
\eqno(4)
$$
where 
$$
H_{\pm g}= \int_0^{\infty} d\omega\,\omega\,a^{\dagger}(\omega)a(\omega)\pm
\int_0^{\infty}d\omega\,\omega \bigl( {\bar g}(\omega) a(\omega)
+ g(\omega) a^{\dagger}(\omega)\bigr) 
\eqno(5)
$$ 
are van Hove Hamiltonians for the bosonic field which are well-known exactly solvable toy 
models of renormalization, both in the infrared and 
ultraviolet regimes. As pure decoherence is a low energy phenomenon we introduce an
ultraviolet cut-off ($g(\omega)= 0$ for $\omega > \omega_c$) and for simplicity we assume
a power-like low energy scaling 
$$
|g(\omega)|^2\sim \omega^{\kappa -1}\ , {\rm for}\ \omega << \omega_c\ .
\eqno(6)
$$
The most frequently used in the context of decoherence is the {\it ohmic coupling} i.e. $\kappa
=0$. 
Under the assumptions of above one obtains the following rigorous results concerning the 
existence and properties of the van Hove Hamiltonians [12]:
\par
1) $\kappa > 0$, regular case, ground states for $H_{\pm g}$ exist,
\par
2) $-1 < \kappa \leq 0$, infrared problem, (includes ohmic case!), $H_{\pm g}$ - bounded
from below but ground states do not exist (in the Fock space),
\par
3) $\kappa\leq -1$, uphysical case, $H_{\pm g}$ - unbounded from below or do not exist
as self-adjoint operators on the Fock space.
\par
The main tool used in the analysis of the Hamiltonian is its diagonalization
in terms of unitary Weyl operators $W(f)=\exp\{a(f)-a^{\dagger}(f)\}$ with  
$\|f\|^2 = \int_0^\infty |f(\omega)|^2 d\omega <\infty\ ,$
acting on the Fock space and satisfying Weyl commutation relations:
$W(f)^{\dagger} = W(-f),\ W(f)W(h) = e^{-i{\rm Im}<f,h>} W(f+h) 
,\ W(f) a(\omega)W(f)^{\dagger} = a(\omega) + f(\omega){\bf 1} $.
The vectors $W(f)\Omega$ are called {\it coherent states} and they form an overcomplete set in
the following sense. If for a given vector $\Psi$ from the bosonic Fock space and any 
$f\in L^2[0,\infty)$ $<\Psi , W(f)\Omega> = 0$, then $\Psi =0$. Taking into
account Weyl relations and $<\Omega ,W(f)\Omega> = \exp\{-(1/2)\|f\|^2\}$ we
obtain
$$
|<W(f)\Omega ,W(h)\Omega>|^2 = e^{-\|f-h\|^2} \leq e^{-(\|f\|-\|h\|)^2}\ ,
\eqno(7)
$$
so $W(f)$ is not unitary on the Fock space unless $\|f\| <\infty$. 
Introducing now for a given formfactor $g(\omega), \|g\| <\infty$ the unitary operator on 
${\cal H}_{SB}$ defined as ${\bf W}(g)={\rm diag} [W(g) , W(-g)]$ we obtain the diagonalized form
$$
{\bf W}(g) {\bf H}_g {\bf W}(g)^{\dagger} = \int_0^{\infty} 
d\omega\,\omega\,a^{\dagger}(\omega)a(\omega) -  E_g
\eqno(8)
$$
where $ E_g = <g,h_1 g> = \int_0^{\infty} \omega |g(\omega)|^2 d\omega$ .
Therefore, the degenerated ground states of ${\bf H}_g$ are given by
$$
{\bf H}_g \Phi_{\pm}(g) = -E_g \Phi_{\pm}(g)\ ,\ \ 
\Phi_{\pm}(g) = \psi_{\pm}\otimes W(\pm g)\Omega\ .
\eqno(9)
$$
Obviously, in our setting the regular case 1)($\kappa > 0$) corresponds to the condition
$\|g\|<\infty$. Two degenerated
ground states of the Hamiltonian ${\bf H}_g$ should be interpreted as the states
of a {\it dressed spin} which consists of a {\it bare spin} and a {\it cloud} of virtual bosons
represented by the coherent states $W(\pm g)\Omega$. The dressed system is decoupled from
the environment and its dynamics is trivial.
\par
The case 2) ($-1 <\kappa \leq 0$) corresponds to $\|g\|=\infty , E_g <\infty$. 
In principle, the states $\Phi_{\pm} (g) $ treated as limits of "normal" states from the Hilbert 
space ${\cal H}_{SB}$ with $\|g\| \to\infty$ can exist in the sense of state functionals on the algebra 
of observables.  But in this case they are {\it disjoint}, i.e. they define
nonequivalent representations of the algebra of observables (van Hove phenomenon [12]). 
Formally, it follows from the
formula $\Phi_+(g) = \sigma_1\otimes W(2g)\Phi_-(g)$ which for $\|g\|\to\infty$ indicates that
there exists no unitary operator which transforms $\Phi_-(g)$ into $\Phi_+(g)$ [13]. 
Physically, it means that
their superpositions are indistinguishable from their mixtures ({\it superselection rule}). 
Some authors 
invoke this mechanism to describe the emergence of classical observables for quantum systems 
[14,16]. This phenomenon should be called {\it static
decoherence} because the disjointness is a permanent feature of these states.
Although from the mathematical point of view 
this is an atractive approach, on the other hand it can lead to profound interpretational 
difficulties, e.g. unphysical superselection rules in quantum electrodynamics[16,17].
\par 
We invoke now  the standard dynamical approach to decoherence in open systems applied
to our spin-boson model. As an initial state of the total system one chooses
the product state 
$$
\Psi_{in} = \psi\otimes\Omega\ ,\ \psi = \alpha_-\psi_- +\alpha_+\psi_+\ ,\   
\alpha_{\pm} \in {\bf C}
\eqno(10)
$$
satisfying
$$
|<\Psi_{in}|\Phi_{\pm}(g)>|^2 = e^{-\|g\|^2}\ ,\ E(\Psi_{in}) = <\Psi_{in}|{\bf H}_g|\Psi_{in}>
=0\ ,
\eqno(11)
$$
computes its time evolution governed by the Hamiltonian (4)(5)
$$
\Psi(t) = e^{-it{\bf H}_g}\Psi_{in}= 
\exp\{i(tE_g - {\rm Im} <g|g_t>)\} \bigl(\alpha_-\psi_-\otimes W(g_t-g)\Omega +
\alpha_+\psi_+ \otimes W(g-g_t)\Omega\bigr)
\eqno(12)
$$
where $g_t(\omega)= e^{-i\omega t}g(\omega)$
and calculates the reduced density matrix for spin 
$$
\rho_t = {\rm Tr}_B |\Psi(t)><\Psi(t)| =
\pmatrix{|\alpha_+|^2 & \alpha_+ \overline{\alpha}_- e^{-\gamma_t}      \cr
          \overline{\alpha}_+ \alpha_- e^{-\gamma_t} & |\alpha_-|^2  \cr}
\eqno(13)
$$
with
$$
\gamma_t = 2\|g-g_t\|^2 \leq 8 \|g\|^2\ .
\eqno(14)
$$
Usually, one discusses the structure of the reduced density matrix (13) only, and identifies
$\gamma_t$ with the decoherence factor. It follows from (14) that decoherence is complete
only if $\|g\|=\infty$ i.e. for a singular coupling. To obtain an asymptotically exponential 
decay of the off-diagonal elements of the
reduced density matrix (13) we must assume 
$$ 
0 < \gamma = \lim_{t\to\infty} {\gamma_t\over t} =
\lim_{t\to\infty} \int_0^{\omega_c}\omega^2|g(\omega)|^2 {1-\cos \omega t\over t\omega^2}
\ d\omega = \pi\lim_{\omega\to 0} \omega^2 |g(\omega)|^2\ .
\eqno(15)
$$
This result agrees with the standard wisdom relating the pure decoherence rate to the value
at $\omega = 0$ of the {\it spectral density function} [4,18] 
$$
{\hat R}(\omega) =  \int_{-\infty}^{\infty}
e^{i\omega t}< R(t)R>_B dt
\eqno(16)
$$
where $R$ is a bath's operator appearing in the interaction Hamiltonian $\sigma_3\otimes R$
and $<\cdot>_B$ is an average with respect to the environment's state. 
It is a special case of the quantum {\it fluctuation - dissipation
theorem} which in fact should be called in this context a "fluctuation-decoherence theorem".
For our model $R = \int d\omega\,\omega ({\bar g}(\omega)a(\omega) + h.c.)$ and hence 
${\hat R}_0(\omega) = 2\pi\omega^2 
|g(\omega)|^2$ where the subscript "$0$" indicates the zero-temperature (vacuum) state
of the bath.
However, a non zero value of $\gamma$ means $\kappa = -1$ which is the uphysical (subohmic) case 2).
The situation is slightly less singular for the temperature $T>0$. It is easy to check that 
for $\omega << T$ (we put $\hbar\equiv k_B \equiv 1$) 
${\hat R}_T(\omega) \simeq (T/\omega){\hat R}_0(\omega)$ and hence the condition of finite
decoherence rate
is satisfied for the  ohmic case $\kappa=0$. The same ohmic assumption is made in the derivation
of the popular Caldeira-Leggett equation for the quantum Brownian particle coupled to
harmonic oscillator heat bath[10].
\par
A different physical interpretation of the discussed process follows from the analysis
of the state evolution for the total system in the regular case. As for $t\to\infty$ the traveling wave
$g_t$ becomes orthogonal to $g$ the asymptotic form of $\Psi(t)$ possesses the structure
of superposition of two triple product states $\psi_{\pm}\otimes W(\pm g)\Omega\otimes 
W(\mp g_t)\Omega$.
Hence, the evolution of the initial product
state (10) given by (12) describes the process of formation of the cloud accompanied by 
emission of the average energy $E_g$ in a form of coherent traveling waves $\pm g_t$. 
Therefore, from the physical point of view the discussed model describes a phenomenon
which should be called {\it false decoherence} [19].
\par
Summarizing the above results one can say that the linear coupling to harmonic oscillator bath
is not an appropriate model of pure decoherence phenomena. In general, we have two different choices of the formfactor $g(\omega)$ - a singular and a regular one, both leading to difficulties. 
\par
The singular coupling (e.g. ohmic,
 $\kappa =0$, for $T>0$ or subohmic, $\kappa=-1$, for $T=0$), which gives a finite asymptotic 
decoherence rate in the formal derivation of the reduced density matrix, leads to serious interpretational difficulties due to infrared divergencies. The ground states do not exist in the Fock space, those defined by limiting procedures describe disjoint states for which superposition principle is not applicable. As a consequence the picture of an open system gradually loosing its "quantum coherences" is not valid in this case.
\par
For the regular coupling the natural representation of dressed states gives the picture of a "physical dressed system" decoupled from its environment. The choice of the initial product state of a bare system and a bath makes sense only in the unique moment of system's creation followed by the irreversible dressing process. After that, the system cannot be prepared in a such state again. 
The absence of true decoherence in the regular (superohmic) case is indicated by the relation
$\lim_{\omega\to 0} {\hat R}(\omega) =0$. 
It follows that the models based on vacuum
fluctuations of the background quantum fields (gravitational, electromagnetic,...) [20] 
are very
unlikely to solve the problem of transition from the quantum to classical world.
\par
{\it Chaotic vs. regular environments}. The similar physical mechanism leading to the absence of pure decoherence appears
in the case of a bath which is an ensemble of quantum subsystems with chaotic properties.
The intuition supported by some heuristic
arguments suggests, as it is formulated in [21], that "...one would expect that 
environments with
unstable dynamics will be much more efficient decoherers,...". However, it follows from the previous discussion that 
the reservoir's energy eigenstates should be degenerated and labeled by other quantum 
numbers which
can be altered without energy modification. On the contrary, for a chaotic system its energy levels
are typically nondegenerated due to the mechanism of {\it level repulsion} [22]. 
\par
We begin with the physical example as a motivation. Consider a large system (say a molecule)
with a relevant collective, single degree of freedom which can be modeled by a 2-level system as above.
This degree of freedom is our open system again, while the internal degrees of freedom of 
the molecule form a bath which can be treated as a large ensemble of quantum systems. 
\par
The simplified mathematical model consists of a $1/2$-spin system interacting with an 
ensemble of $N$ identical $M$-level quantum systems by means of the following mean-field type 
Hamiltonian which is an anolog of (2)[23]
$$
{\bf H}_Q = {1\over 2}\epsilon\sigma_1 + \sum_{k=1}^N h^{(k)} + \sigma_3\otimes N^{-1/2}\sum _{k=1}^N Q^{(k)} \ .
\eqno(17)
$$
Here $h^{(k)}$ is a copy of the Hamiltonian with the spectral resolution
$h=\sum_{m=1}^M \epsilon_m |m><m|$, $\epsilon_{m+1} \geq \epsilon_m$ 
and $Q^{(k)}$ is a copy of an operator $Q=Q^{\dagger}$, ${\rm Tr}Q =0$. The rerefence
state of environment is assumed to be the product state $\otimes_{k=1}^N \rho^{(k)}$
where $\rho^{(k)}$ is a copy of the microcanonical state giving an uniform probability 
distribution over all states $|m>$. We assume again, that the tunneling time $\hbar/\epsilon$
is much longer that the decoherence time and that the energy $\epsilon$ is much smaller
than the typical reservoir's constituents energy spacing $\Delta$ (compare again [8]). 
For $N\to\infty$
the mean-field reservoir's observable $N^{-1/2}\sum_{k=1}^N Q^{(k)}$ behaves like a
Gaussian noise and in the Markovian approximation the pure decoherence rate $\gamma$ 
for the spin is given by the following version of the fluctuation-dissipation formula
$$
\gamma = {1\over 2}\lim_{\omega\to 0}{\hat R}(\omega)\ ,\ {\hat R}(\omega) =
{\pi\over M}\sum_{m,m'=1}^M |<m|Q|m'>|^2 \delta \bigl((\epsilon_m - 
\epsilon_{m'})-\omega\bigr)\ .
\eqno(18)
$$ 
This formula makes sense also when instead of identical subsystems the reservoir consists 
of a large random ensemble of quantum systems with Hamiltonians $h^{(k)}$ characterized by the 
average nearest-neighbour level spacing distribution $p(s)$ with the average $\Delta$. For $\omega << \Delta$ only the nearest-neighbour level spacings $s=\epsilon_{m+1} - \epsilon_m$
contribute to the spectral function ${\hat R}(\omega)$ (18)and therefore the difference between
the bath consisting of classically integrable or chaotic systems becomes crucial. For the former we expect
a Poisson distribution of $s$ while for the later the level repulsion given by 
$p(s)\sim s^{\beta} , \beta = 1,2,$ or 4  is generically observed [22].
Assuming that the magnitude of the matrix
elements $|<m+1|Q|m>|$ is not strongly correlated with $\epsilon_{m+1}-\epsilon_m$
we obtain
$$
{\hat R}(\omega)\simeq \pi {\bar Q}^2 p(\omega)
\eqno(19)
$$
where ${\bar Q}^2$ is an averaged value of $|<m+1|Q|m>|^2$. As a consequence 
pure decoherence rate is equal to zero for the chaotic systems while for the regular ones we obtain a finite value of $\gamma$. 
\par
We have shown that for important models used to analyse environmental decoherence this
phenomenon disappears in the limit of pure decoherence. These cases can be easily detected applying
the unified approach of spectral density function (16) at $\omega\to 0$.To avoid the possible
influence of approximation procedures we have studied in details an exactly solvable spin-boson model which illustrated both , vanishing pure decoherence and infrared divergencies for the singular choice of the coupling.
The proper models of pure decoherence
involve {\it elastic
scattering} processes which "for all practical purposes "
provide strong enough suppression of quantum superpositions in the macroworld
(e.g. Joos and Zeh [7]),[24] or describe quantum Brownian motion [25].

\acknowledgments
The author thanks Phillipe Blanchard, Mark Fannes, Fritz Haake, Micha\l\ , Pawe\l\ and
Ryszard Horodecki, 
Robert Olkiewicz and Karol \.Zyczkowski  for discussions. The work is partially supported by the
KBN Grant BW/5400-5-0255-3.


\begin{references}
\bibitem{[8]} As  well-known physical examples one can take two molecules with left-right
symmetry: a small one - amonium- with the tunelling time $\sim 10^{-10}$ sec., and a big one 
 -alanine- with the estimated tunneling time longer than $10^9$ years [14].
\bibitem{[13]} In the singular case the very definition of ground states is not unique and depends on the limiting procedure and the choice of algebra of physical observables. The detailed rigorous analysis of ground states defined as limits of the thermal states for $T\to 0$ and for spin-boson model with $\epsilon >0$ shows a phenomenon of spontaneous symmetry breaking. Namely,
for the ohmic case ($\lim_{\omega\to 0}\omega|g(\omega)|^2 = \alpha$)  if the coupling constant
$\alpha$ is larger than a certain critical value $\alpha_c \approx 1$ (for small $\epsilon$) 
we obtain two disjoint ground states (left and right) while for $\alpha < \alpha_c$ the ground state is unique [15]. Qualitatively, the similar behaviour was already predicted in [11].
\end{references}
\end{document}